\documentclass[%
 aip,
 amsmath,amssymb,
 reprint,%
 ]{revtex4-2}
 \usepackage[utf8]{inputenc}
 \usepackage{color}
\usepackage[draft]{changes}
\usepackage{ulem}
\usepackage{amsfonts,amsmath,bm}
\usepackage{natbib}
\usepackage{graphicx}

\begin{document}

\title{Unveiling the connectivity of complex networks using ordinal transition methods}

\author{Juan A. Almendral}
\email{juan.almendral@urjc.es}
\affiliation{Complex Systems Group \& GISC, Universidad Rey Juan Carlos,
28933 M\'ostoles, Madrid, Spain}

\affiliation{Center for Biomedical Technology, Universidad Polit\'ecnica de
Madrid, 28223 Pozuelo de Alarc\'on, Madrid, Spain}

\author{I. Leyva}

\author{Irene Sendi\~na-Nadal}

\date{\today}

\begin{abstract}
  Ordinal measures provide a valuable collection of tools for analyzing correlated data series. However, using these methods to understand the information interchange in networks of dynamical systems, and uncover the interplay between dynamics and structure during the synchronization process, remains relatively unexplored. Here, we compare the ordinal permutation entropy, a standard complexity measure in the literature, and the permutation entropy of the ordinal transition probability matrix that describes the transitions between the ordinal patterns derived from a time series. We find that the permutation entropy based on the ordinal transition matrix outperforms the rest of the tested measures in discriminating the topological role of networked chaotic R\"ossler systems. Since the method is based on permutation entropy measures, it can be  applied to arbitrary real-world time series exhibiting correlations originating from an existing underlying unknown network structure. In particular, we show the effectiveness of our method using experimental datasets of networks of nonlinear oscillators.
  \end{abstract}

\maketitle


\section{Introduction}

Time series analysis has garnered significant research attention in recent decades. However, the exponential growth in data generation from various social, technological, and natural sources observed in the last years has posed a challenge for researchers seeking to extract valuable information from these datasets. Among the array of new tools developed for this purpose, the ordinal methods derived from the seminal work of Bandt and Pompe \cite{Bandt2002} have emerged as particularly intriguing.

In this approach, the original data series undergoes a process of coarse-graining, wherein it is replaced by a reduced set of symbols representing the order permutations of consecutive data points. Statistical properties and correlations of those ordinal permutation series effectively capture much of the dynamical information inherent to the original system. Moreover,  its analysis is faster, computationally affordable and more robust to noise than raw data analysis. As a result, the applications of ordinal methods continue to expand \cite{Leyva2022}, encompassing diverse fields such as neuronal \cite{Tlaie2019b} and brain dynamics \cite{Lehnertz2023}, laser dynamics \cite{Tiana2019}, and sports data analysis \cite{buldu2021}, among others.

Recently, the field of ordinal methods has advanced to incorporate "ordinal transition networks" (OTN). Initially proposed in Ref. \cite{Small2013}, this concept introduces an additional layer of temporal correlation to the analysis by examining the statistics of ordinal patterns and their transitions. The time series is now represented as a network, where each ordinal pattern corresponds to a node, and the possible transitions among them are the links.

This innovative tool has demonstrated its potential in detecting subtle dynamical changes \cite{Sun2014, McCullough2015}, and its associated ordinal transition entropy has proven to be more robust than standard permutation entropy when dealing with noisy signals \cite{Pessa2019, Olivares2020}. Furthermore, the statistics of self-transitions within an OTN \cite{Borges2019, Cardoso-Pereira2022} offers an effective means of characterizing diverse time series dynamics. Notably, OTN complexity can accurately reproduce the results of Lyapunov exponents even for small embedding sizes \cite{Huang2021}.

The versatility of this method extends to various applications, such as distinguishing between different consciousness states \cite{Varley2021}, analyzing EEG data \cite{Kulp2016}, investigating stock markets \cite{Wang2022, Peng2022}, and examining transportation data \cite{Cardoso-Pereira2022}. In combination with complex network techniques for nonlinear time series analysis, such as visibility or recurrence networks \cite{Nicolis2005, Zhang2006, Lacasa2008, Pessa2019, Zou2019, Silva2021}, this approach presents a valuable addition to the set of available tools.

Ordinal methods offer good potential for various applications, particularly in finding correlations between time series. The multivariate extension of these methods enables the synthesis of information from multiple data sources, resulting in a unified set of symbols \cite{Zhang2017, Shahriari2020, Subramaniyam2021}. This approach proves useful in detecting phase transitions within the collective state of small groups of coupled chaotic nodes. Furthermore, by incorporating delays into the analysis, multivariate ordinal methods can unveil the directionality of the coupling relationships \cite{Ruan2019}.

Recent developments have introduced the concept of an ordinal network based on pattern co-occurrence between time series \cite{Ren2020}. This approach facilitates the inference of correlations between different time series. Moreover, the notion of {\it ordinal synchronization} \cite{Echegoyen2019} demonstrates the capability to detect phase and anti-phase synchronization even in noisy real data.

These examples highlight the significant potential of ordinal methods in studying dynamical ensembles and networks. However, it is important to note that most of these applications currently remain confined to proof-of-concept studies involving small networks \cite{Zhang2017, Shahriari2020}. In addition, many of these approaches rely on multivariate pairwise correlations to extract information \cite{Lehnertz2023}.

Nevertheless, it is crucial to realize that each element within a networked ensemble undergoes an information flux that alters its dynamics, effectively encoding valuable information regarding its topological role and the collective state. Ordinal methods serve as an ideal tool for unveiling these dynamical changes, enabling the creation of centrality rankings for nodes without solely relying on pairwise correlations \cite{Tlaie2019, Letellier2020}.

Building upon this premise, our work extends the application of these methods to analyze the synchronization process in complex networks. Our findings demonstrate that ordinal transition methods outperform conventional ordinal patterns' statistics when it comes to detecting subtle dynamical changes and discriminating nodes based on their topological roles. These initial results, using synthetic networks of chaotic R\"ossler systems and data from experiments with nonlinear electronic circuits, illuminate new possibilities for using ordinal methods in various applications, including functional brain data analysis \cite{Lehnertz2023}, power grids, mobility networks, or any other domains involving the close interplay between structural and functional relationships within large-scale dynamical ensembles.

\section{Model and methods}

\subsection{Model}
We consider a network of $N$ identical R\"ossler dynamical systems \cite{Ros76} whose dynamics are governed by the following equations:
\begin{equation}
    \dot{\bf x}_i={\bf f}({\bf x}_i) - \Tilde{\sigma} \sum {\cal L}_{ij}{\bf h}({\bf x}_j), 
    \label{eq:dyn}
\end{equation}
with $i=1,\dots,N$; ${\bf x}_i=(x_i,y_i,z_i)$ the vector state of the node $i$; $\bf{f} \left( \displaystyle \bf{x} \right)$ and $\bf{h} \left( \displaystyle \bf{x} \right): \mathbb{R}^3 \rightarrow \mathbb{R}^3$, being ${\bf f}({\bf x})=[-y-z,x+ay,b+z(x-c)]$ the vector flow of the R\"ossler system, and ${\bf h}({\bf x})=[0,y,0]^T$ the coupling function. We set $a=b=0.2$ and $c=9.0$ to get a phase-coherent chaotic attractor. The coefficients ${\cal L}_{ij}=k_i\delta_{ij}- a_{ij}$ are the elements of the Laplacian matrix whose adjacency matrix ${ A}:=(a_{ij})$ encodes the connectivity among the nodes of the network: $a_{ij}=1$, if  $i$ and $j$ are connected, and $a_{ij}=0$ otherwise. Thus the degree of node $i$ is $k_i=\sum_j a_{ij}$. The constant $\Tilde{\sigma}=\frac{\sigma}{k_{\rm max}}$ is the coupling strength normalized by the maximum degree present in the network, that is, $k_{\rm max}=\max (k_i)$. This normalization is introduced to properly compare observables between different network realizations \cite{Gomez2007}. The system of $N$ equations described by (\ref{eq:dyn}) has been numerically integrated using a Runge-Kutta method of 4th order with a time discretization of $0.005$. In all simulations, the time evolution is extended up to $12,000$ time units, discarding the first half, which is considered a transient.

In ordinal methods, how the raw data is projected into an ordinal series depends on the particularities of the data, their sampling, or their continuous or discrete nature, without affecting the rest of the procedure \cite{McCullough2015}. In our case, to extract information about the temporal organization of each nodal dynamics ${\bf x}_i(t)$, we first computed the two-dimensional Poincar\'e section ${\cal P}\equiv \{[x_i(t_m),z_i(t_m)]\in \mathbb{R}^2| \, \dot{y}_i(t_m)=0,\ddot{y}_i(t_m)>0\}$ \cite{Shahriari2023}. This allows us to map the whole attractor ${\bf x}_i$ of node $i$ into the one-dimensional time series ${\cal S}_i \equiv \{ y_i(t_m),m=1,\dots,M\}$, generated at the times $t_m$ the attractor crosses the section ${\cal P}$. Then, we construct the order relations of $D$ successive data points in the sampled time series ${\cal S}_i$ in the following manner. Once the terms in the sequence ${\cal S}_i$ are split into disjoint blocks of size $D$, we create a symbolic sequence in which each element is replaced by a number in $[1,\dots,D]$, corresponding to its relative ranking respect to its $D-1$ neighbours in the block. Therefore, each block is mapped into one of the $D!$ possible permutations in which $D$ different elements can be arranged. We refer to these permutations as ordinal patterns, using the notation $\pi_{\ell}$ with $\ell = 1, \dots, D!$. As an example, let us consider the series $\{2.3, 3.4, -2.7, 0.4, 1.6, 2.9, -2.8, -0.5, 3.1, 2.4, \ldots\}$. We first split the series into disjoint blocks of size $D=3$: $\{2.3,3.4,-2.7\}$, $\{0.4,1.6,2.9\}$, $\{3.1,-0.5,3.8\}$, $\{2.4, \ldots\}$. Then, we derive the ordinal pattern for each block. It can be done from maximum to minimum or the other way around. In the first case, the ordinal patterns would be $\{2,1,3\}$, $\{3,2,1\}$, $\{2,3,1\}$, $\ldots$, which  are arbitrarily denoted as  $\pi_5$, $\pi_1$, $\pi_2$, $\ldots$ (see Fig.~\ref{fig2} for our notation of the six possible permutations).

Finally, we define the probability of occurrence of a given pattern $\pi_{\ell}$ as $p_{\ell}=\#(\pi_{\ell})/L$, being $\#(\pi_{\ell})$ the number of times the ordinal pattern $\pi_{\ell}$ appears in the sequence ${\cal S}_i$ and $L= \lfloor M/D \rfloor$ the total number of blocks of size $D$ in which we divide the series ${\cal S}_i$ ($\lfloor \, \rfloor$ is the floor function). Note that this procedure is only meaningful if $M \gg D!$.

\subsection{Methods}

In this Section, we present the methods employed to characterize the statistical complexity of a nodal dynamics. Our ultimate objective is to establish a relationship between the dynamical behaviour of each node and its structural connectivity within the network. To achieve this, we compare the ordinal permutation entropy based on the probability distribution of ordinal patterns and the ordinal transition entropy based on the transition probabilities between consecutive non-overlapping ordinal patterns.

Permutation entropy has previously been identified as a reliable indicator of the topological role of a node within a dynamical network \cite{Tlaie2019, Letellier2020}. However, our study reveals that analysing the transition probabilities between ordinal patterns offers a more effective and informative measure for assessing a node's degree centrality.

\subsubsection{Ordinal permutation entropy}

Given the probability distribution of the ordinal patterns $\pi_{\ell}$ of size $D$, with $\ell=1,\dots,D!$, we define the normalized permutation entropy as the Shannon entropy evaluated on the ordinal pattern probability distribution: 
\begin{equation}\label{eqH}
    {\cal H}_0 =-\frac{1}{\ln D!}\sum_{\ell} p_{\ell}\ln{p_{\ell}},
\end{equation}
with the criterion $0^0=1$ to deal the case $p_{\ell} =0$. According to Bandt and Pompe \cite{Bandt2002}, $3\le D\le 7$ values provide reliable information on the natural complexity of time series coming from chaotic dynamical systems as long as $M\gg D!$. However,  unobserved ordinal patterns have been reported in chaotic dynamical systems, no matter how large the time series is, due to the underlying temporal correlations \cite{Amigo2007}.  

\begin{figure*}
    \centering   \includegraphics[width=0.7\textwidth]{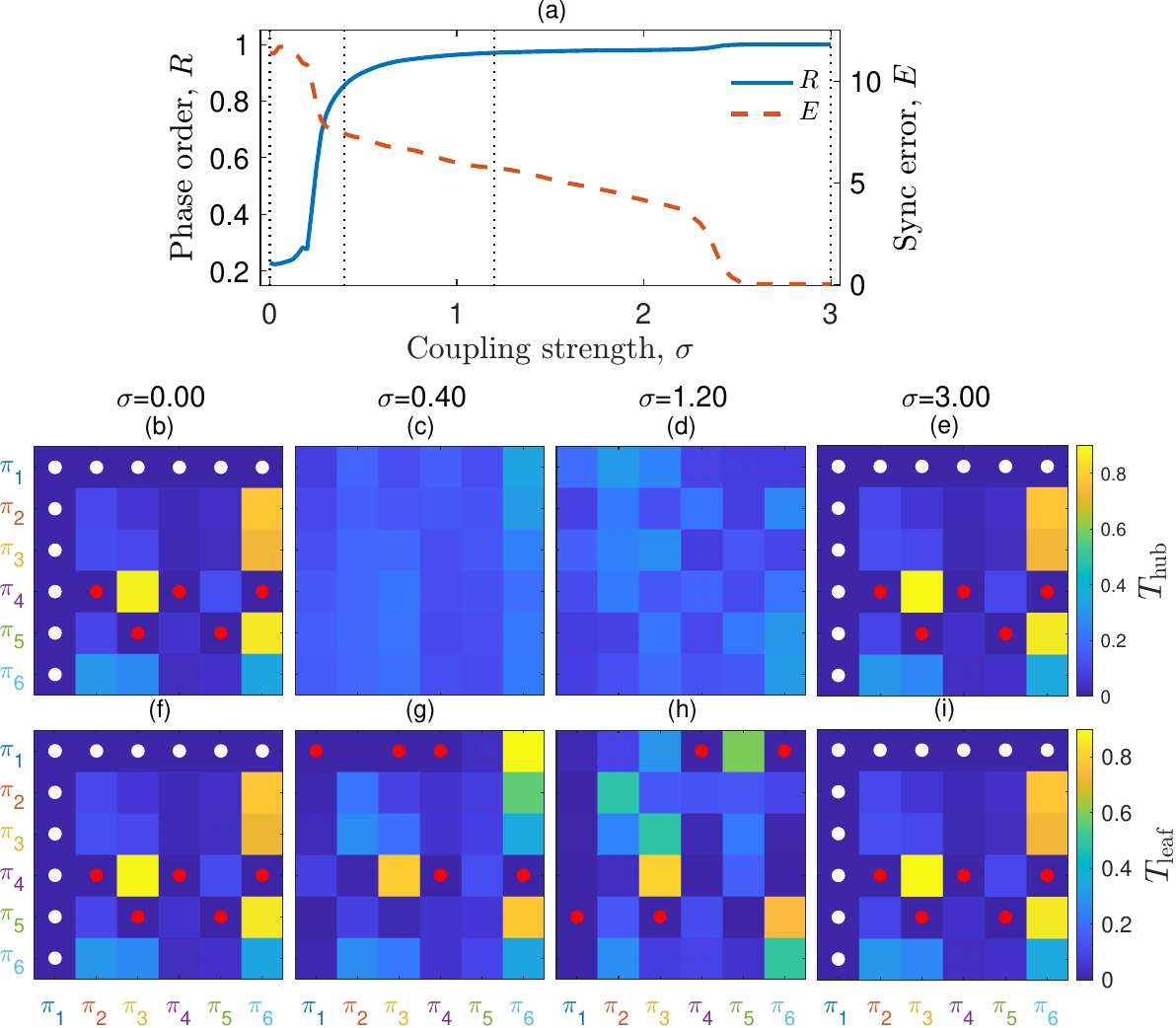}
    \caption{OTP matrices for a star graph of $N=16$ identical R\"ossler systems along the route to synchronization. (Top panel) Phase order $R$ (left axis) and synchronization error $E$ (right axis) as a function of the coupling strength $\sigma$. (Color panels) OTP matrices $T$, with ordinal patterns $\pi_\ell$ ($\ell={1,\dots,6}$ since $D=3$), of the hub (top row) and one of the leaves (bottom row), for the coupling values marked with dotted lines in the top panel along the synchronization process. Ordinal transitions with zero probability are marked with dots: in white, those caused for $\pi_1$ being a forbidden ordinal pattern, and in red, the transitions that, despite being between existing ordinal patterns, they actually do not occur. Time series length $M=1000$. R\"ossler parameters: $a=b=0.2$ and $c=9.0$.}
    \label{fig1}
  \end{figure*}
  
\subsubsection{Ordinal transition entropy}

In addition to the probability $p_{\ell}$ of each ordinal pattern $\pi_{\ell}$, the transition probability $p_{\ell m}$, from the ordinal pattern $\pi_\ell$ to $\pi_m$, may reveal information into the finer temporal organization of a dynamical system \cite{Small2013}. We define the ordinal transition probability (OTP) matrix $T:=(p_{\ell m} )$ as
\begin{equation}\label{transmat}
    p_{\ell m}=\frac{\#(\pi_{\ell},\pi_m)}{\#(\pi_{\ell})}
\end{equation}
being $\#(\pi_{\ell},\pi_m)$ the number of times the pair $\pi_{\ell}-\pi_m$ consecutively occurs in the time series. Note that, in case $\#(\pi_{\ell})=0$ for some pattern $\pi_{\ell}$, we can define $p_{\ell m}=0$. The total number of blocks $L$ of size $D$ must now be $L\gg D!^2$ so that the OTP matrix $T$ is statistically significant.

Equation (\ref{transmat}) is a proper stochastic matrix whose weights encode an OTN among ordinal patterns, including self-transitions, into which the time series of each nodal dynamics can be mapped. Hence, the complexity of this OTN will depend on the  diversity of both ordinal patterns and transitions occurring among them.

Since $\sum_m p_{\ell m}=1$, we can define the node permutation entropy ${\cal H}_{\pi_{\ell}}$ associated with the ordinal pattern $\pi_\ell$, a node of the OTN, which quantifies the randomness of the local transitions from the ordinal pattern $\pi_{\ell}$ to any other pattern \cite{Masoller2015,McCullough2017}, as
\begin{equation}\label{eqHpi}
    {\cal H}_{\pi_{\ell}}=-\frac{1}{ln D!}\sum_{m_=1}^{D!} p_{\ell m}\ln p_{\ell m}.
\end{equation}

We characterize the transitional complexity of the OTN at the global level with a {\em network} permutation entropy obtained as the average of the node permutation entropies given by Eq.~(\ref{eqHpi}). Depending on how the average is performed, we consider using either the first moment of the distribution of the ${\cal H}_{\pi{_{\ell}}}$ values as in \cite{Masoller2015}
\begin{equation}\label{eqHT}
   {\cal H}_{\rm T} = 
   \frac{1}{D!}\sum_{\ell =1}^{D!}   {\cal H}_{\pi_{\ell}}   
\end{equation}
or, alternatively,  as defined in \cite{Unakafov2014}:
\begin{equation}\label{eqHTz}
   {\cal {\hat H}}_{\rm T} = 
   \sum_{\ell =1}^{D!}  p_{\ell} {\cal H}_{\pi_l}
\end{equation}
which characterizes the weighted average (over the stationary probabilities $p_{\ell}$ of each pattern $\pi_{\ell}$) of the diversity of consecutive ordinal patterns. 
Other measures to characterize ordinal transition networks can be found in Refs.~\cite{Zou2019,Zanin21}.

\subsubsection{Synchronization measures}

In addition to characterizing the nodal dynamics by the randomness of the ordinal patterns and their transitions, we evaluate  the dynamical network's collective state  for increasing coupling values since the chosen networked system (\ref{eq:dyn}) is known to evolve from a totally incoherent state when $\sigma=0$ to a regime where the phases are locked while the amplitudes vary chaotically and uncorrelated, up to a regime of complete synchronization for very large $\sigma$. We compute the time-averaged  phase order parameter
\begin{equation}
R = \frac{1}{N} \langle | \sum_{j=1}^N {\rm e}^{{{\rm i } \theta_j}} | \rangle_t    
\end{equation}
with the phase $\theta_j$ of the $j$-oscillator defined as $\theta_j = \arctan (y_j/x_j)$ \cite{Boc2002}, and synchronization error
\begin{equation}
E = \frac{2}{N(N-1)} \langle \sum_{i \neq j} \| \mathbf{x}_i-\mathbf{x}_j \| \rangle_t,
\end{equation}
which account for the level of phase ($0\le R\le 1$) and total synchronization ($E\ge 0$) respectively. When the network is in complete synchrony, $R=1$ and $E=0$. Here, $\langle \rangle_t$ stands for the time average along a sufficiently large time series.
\begin{figure*}[t]
    \centering   \includegraphics[width=\textwidth]{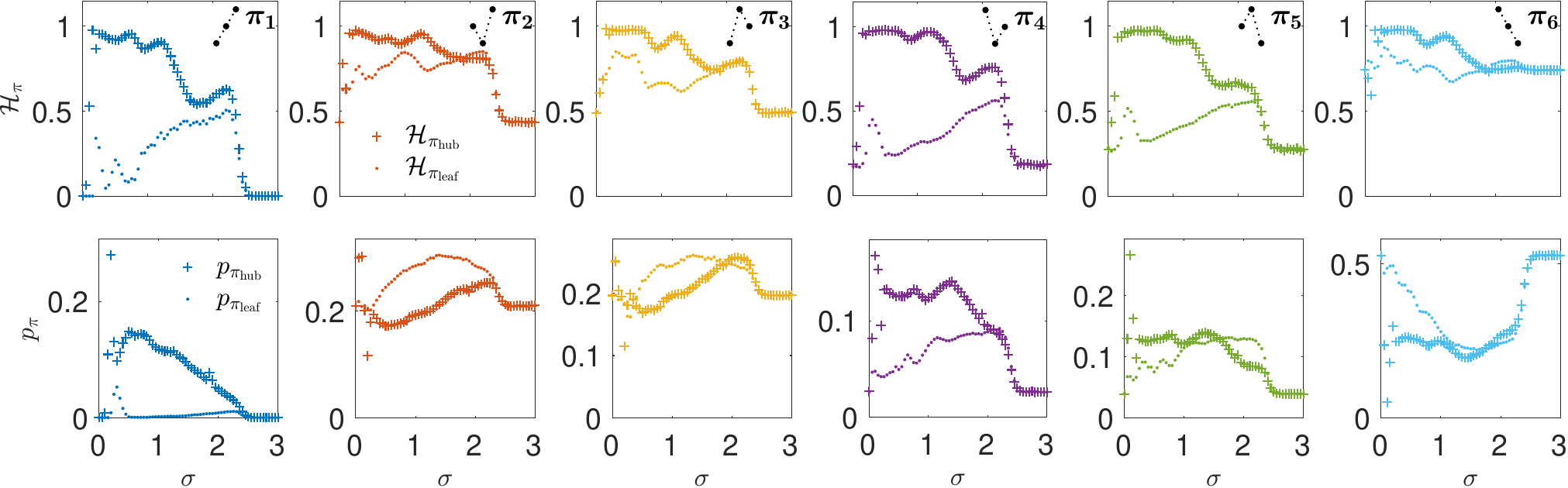}
    \caption{Evolution of the node permutation entropies ${\cal H}_{\pi_\ell}$ (top row) and the corresponding probabilities $p_{\pi_\ell}$ (bottom row) of each ordinal pattern $\pi_\ell$, for $D=3$, as a function of the coupling $\sigma$ (crosses for the hub and dots for one of the leaves). Notice that the scale is not the same for all the panels at the bottom row. In the upper right corner of the panels in the top row, it is shown schematically with three black dots the permutation of the corresponding ordinal pattern (for instance, $\pi_2$ is $312$ and $\pi_3$ is $231$). Data is for the same $N=16$ star of R\"ossler systems and parameters as in Fig.~\ref{fig1}.
    \label{fig2}}
\end{figure*}

\section{Results}

\subsection{Star network}
Let us start with a star configuration of $N$ coupled R\"ossler systems. This network topology has $N-1$ nodes of degree $k_{\rm leaf}=1$ connected to a central one, the hub, with $k_{\rm hub}=N-1$, thus, offering two types of nodal dynamics with the maximum topological distance possible. 


Figure \ref{fig1} illustrates the transition to synchronization of a $N=9$ star as the coupling strength $\sigma$ increases. Along this route, the initially identical dynamics exhibited by the hub and the leaves start to differentiate due to the coupling interaction. This differentiation becomes evident when examining the OTP matrix $T$ for $D=3$, which has six ordinal patterns (corresponding to the following permutations: $\pi_1 \equiv 321$, $\pi_2 \equiv 312$, $\pi_3 \equiv 231$, $\pi_4 \equiv 132$, $\pi_5 \equiv 213$, $\pi_6 \equiv 123$).

The colormap panels depict the OTP matrices $T$ of the hub (b-e) and one of the leaves (f-i), representing four different values of $\sigma$ corresponding to various synchronization stages. When $\sigma=0$ (panels b, f) and $\sigma=0.2$ (panels e, i), the hub and the leaves' OTP matrices exhibit the same color coding. This similarity arises because they describe the transition probabilities between ordinal patterns of the same intrinsic dynamics, given by the flow $\mathbf{f}$ in Eq.~(\ref{eq:dyn}) when the systems are uncoupled or coupled but synchronized. In these panels, white and red dots indicate unobserved ordinal transitions ($p_{\ell m}=0$). These transitions may be absent either because the chosen chaotic dynamics include one forbidden ordinal pattern, white dots (caused by the forbidden pattern $\pi_1$), or because unreachable ordinal patterns exist from certain initial states, red dots (for instance, a $\pi_3$ pattern cannot follow a $\pi_5$ pattern).

As soon as the hub and the leaf interact (panels c, g and d, h), the colormap changes differently for each of them. New transitions appear while others disappear. Notably, all ordinal transitions become nearly equiprobable for the hub, which is indicative of noisy dynamics—a characteristic feature.

To closely inspect how those transitions between ordinal patterns evolve along the synchronization process for each type of node in a star graph, we plot in each panel of the top row of Fig.~\ref{fig2} the node permutation entropy ${\cal H}_{\pi_\ell}$ of each ordinal pattern $\pi_\ell$ for the hub and one of the leaves and  the corresponding ordinal pattern frequencies $p_{\pi_{\ell}}$ at the bottom row  as a function of the coupling strength $\sigma$. The most remarkable differences between hub and leaf come from those transitions starting at patterns $\pi_1$, $\pi_4$, and $\pi_5$, since the gap between the node permutation entropies ${\cal H}_{\pi}$ between hub and leaf is the largest, while for the rest of patterns is less pronounced. In particular, the pattern $\pi_1$, which is forbidden in the isolated dynamics, not only emerges due to the interaction but also becomes much more entropic in the hub's dynamics than in the leaves. In addition, note the differences in the probability frequency $p_{\pi}$ of each pattern that will have an effect on the network permutation entropy of the OTN as defined in Eq.~(\ref{eqHTz}).

\begin{figure*}
    \centering   
 \includegraphics[width=0.8\textwidth]{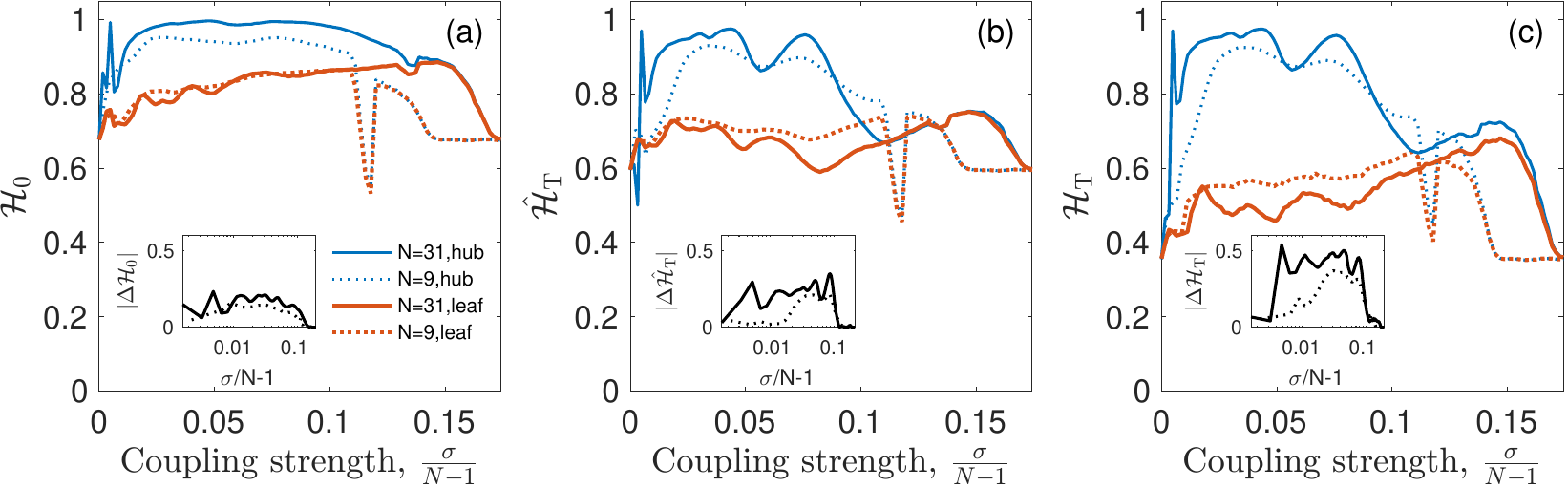}
    \caption{Comparison between (a) the ordinal permutation entropy ${\cal H}_0$, and (b) the weighted ${\hat{\cal H}}_{\rm T}$ and (c)  unweighted ${\cal H_{\rm T}}$ network permutation entropies along the route to synchronization, for the hub (blue crosses) and one of the leaves (red dots), of a star graph of size $N=9$ (dotted curves) and $N=31$ (solid curves). Insets show, in a log-linear scale, the absolute difference between the entropies of hub and leaf ($\Delta {\cal H}_0 = {\cal H}_{0_{\rm hub}} - {\cal H}_{0_{\rm leaf}}$, and the same for $\Delta {\hat{\cal H}}_{\rm T}$ and $\Delta {{\cal H}}_{\rm T}$). 
    \label{fig3}}
\end{figure*}

The primary objective of this work is to evaluate whether an entropic measure based on the information encoded in the OTN can outperform the predictive power of the entropic quantifiers based on just the probability distribution of the ordinal patterns. To examine this, we compare in Fig.~\ref{fig3} how the ordinal permutation entropy ${\cal H}_0$ [Eq.~(\ref{eqH})]  and the network permutation entropies ${\cal H}_{\rm T}$ [Eq.~(\ref{eqHT})] and ${\cal \hat{H}}_{\rm T}$ [Eq.~(\ref{eqHTz})] differentiate between the hub and leaf dynamics for two star networks of $N$=9 and $N$=31 nodes. The network permutation entropy ${\cal H}_{\rm T}$ (panel (b)) effectively separates the hub and leaf dynamics right from the onset of phase synchronization, and maintains this distinction over a broader range of coupling strengths compared to the two other entropies. This is linked to the results shown in Fig.~\ref{fig2}, in which those patterns with the greatest differences between hub and leaf node permutation entropies ($\pi_1$, $\pi_4$ and $\pi_5$) are those for which the probabilities of occurrence are smaller than for the rest ($\pi_2$, $\pi_3$, and $\pi_6$). Note that the scale is not the same for all the panels. Consequently, the weighted version of the network permutation entropy, $\hat{\mathcal{H}}_{\rm T}$, is biased by the most frequent patterns $\pi_2$, $\pi_3$, and $\pi_6$ which are the ones with the most similar hub and leaf node permutation entropies and there, less sensitive to distinguish between the nodes' different roles in the collective dynamics.

Furthermore, a noteworthy observation is that the differentiation in the ${\cal H}_T$ of the hub is more pronounced, and occurs at a lower coupling strength, in the case of the larger star with $N=31$, compared to the smaller one with $N=9$. On the other hand, the values for the leaf nodes in both stars are similar, which is expected as they have the same degree, $k_{\rm leaf}=1$. This finding implies that the network permutation entropy ${\cal H}_T$ has the potential for effectively discerning topological roles within more complex ensembles, as we will explore in the next Section.

\subsection{Scale-free network}

Once we have evidence that the network permutation entropy ${\cal H}_T$ can uncover the information stored in the OTN and discriminate the different roles that nodes have in star networks, we move forward to test this measure in the more challenging task of analyzing the synchronisation process of a scale-free network. Precisely, we consider the network dynamics of $N=300$ nodes, as described by Eq.~(\ref{eq:dyn}), whose connectivity follows a scale-free degree distribution \cite{bar99}.

Given a coupling value $\sigma$, for each node $i$ we compute the corresponding ordinal permutation entropy ${\cal H}^{(i)}_0$ and the network permutation entropy ${\cal H}^{(i)}_{\rm T}$. Since we expect that the nodes with the same degree $k$ will have the same dynamical role within the network, we define a $k$-class average for the network permutation entropies as \cite{Tlaie2019}:
\begin{equation}
\langle {\cal H}_{\rm T} \rangle_k = \frac{1}{N_k} \sum_{ \{ i|k_i=k \} } {\cal H}^{ (i)}_{\rm T},
\end{equation}
where $N_k$ is the number of nodes with  degree $k$ and $\langle \rangle_k$ is just to denote how the  measure has been obtained as an ensemble average of the given measure at the node level restricted to nodes with the same connectivity $k$. Similarly, we define a $k$-class average for the ordinal permutation entropies of those nodes with the same degree: $\langle {\cal H}_0 \rangle_k$.

\begin{figure*}
    \centering   \includegraphics[width=0.7\textwidth]{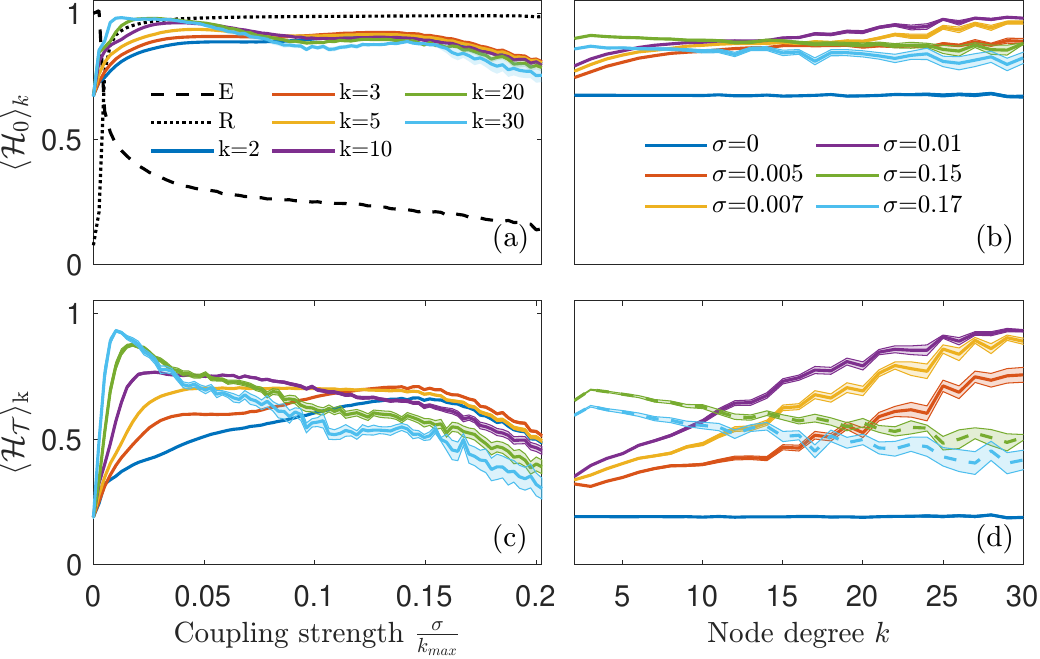}
    \caption{Comparison between the $k$-class ordinal permutation entropy $\langle {\cal H}_0 \rangle_k$ (a,c) and the $k$-class network permutation entropy $\langle {\cal H}_{\rm T} \rangle_k$ (b,d) for heterogeneous scale-free networks of $N=300$ nodes and mean degree $4$: (a,b) as a function of the normalized coupling strength $\frac{\sigma}{k_{max}}$, for several values of degree $k$ class; (c,d) as a function of  $k$, for several values of $\sigma$. In panel (d), solid lines refer to weak coupling values while dashed ones refer to couplings favoring a state close to synchronization. The synchronization error $E$ (rescaled for clarity) and the Kuramoto parameter $R$ have been added in panel (a) as a reference. Each curve is the result of averaging over 50 network instances. Shaded bands indicate the confidence interval around the mean value computed as three times the standard error of the mean.}
    \label{fig4}
\end{figure*}

The results are presented in Fig.\ref{fig4}, which compares $\langle {\cal H}_0 \rangle_k$ (a,c) and $\langle {\cal H}_{\rm T} \rangle_k$ (b,d). It is clear that the network permutation entropy surpasses the ordinal permutation entropy in its ability to sort nodes according to their degree. Upon increasing the normalized coupling strength $\sigma/k_{\text{max}}$, both entropies exhibit a distinct separation based on node degrees. However, the differences between $k$-classes are more pronounced in the case of $\langle {\cal H}_T \rangle_k$.

It is worth noting that, for weak values of the coupling network, hubs exhibit higher entropy values, similar to the behavior of the central node of a star network. However, as the synchronization progresses further, the ranking of the degree classes reverses. This change in behaviour throughout the synchronization process reflects an interesting fact: in weakly coupled networks, highly connected nodes perceive the information from the network as a source of noise, thereby increasing their entropy above that of the low-connected nodes. However, beyond this point, the highly connected nodes take the lead in driving the transition to coherence, while the other nodes remain unsynchronized \cite{Zhou2006, Pereira2010}, resulting, as a consequence, in an exchange of entropy trends.

The results shown in Figs.~\ref{fig4}(b,d) shed light on understanding this entropy-based centrality ranking. We plot the ordinal permutation entropy (b) and the network permutation entropy (d) as a function of $k$ for various coupling values. The entropies of the degree classes demonstrate a quasi-linear relationship with $k$, displaying a positive slope for weak coupling (solid lines) and a negative slope for coupling strengths close to the system's  synchronization (dashed lines). Therefore, network permutation entropy measures stand out as a superior choice. Consequently, a centrality ranking can be established solely based on this entropy without prior knowledge of the underlying structure or costly pairwise computations of the observed time series.

To assess the method's validity in a more realistic environment with available ground truth structural information, we analysed the experimental datasets of networks of  nonlinear electronic circuits provided by Ref. \cite{Vera2020}. These datasets comprise the time series of the output voltage of $N=28$ electronic circuits coupled in 20 different network configurations and monitored  along their synchronization process for 100 coupling levels, ranging from disconnection (isolated nodes) to values producing a network state of complete synchrony. Please refer to the reference \cite{Vera2020} for a full description of the experiments.

\begin{figure}[thb]
    \centering   \includegraphics[width=0.5\textwidth]{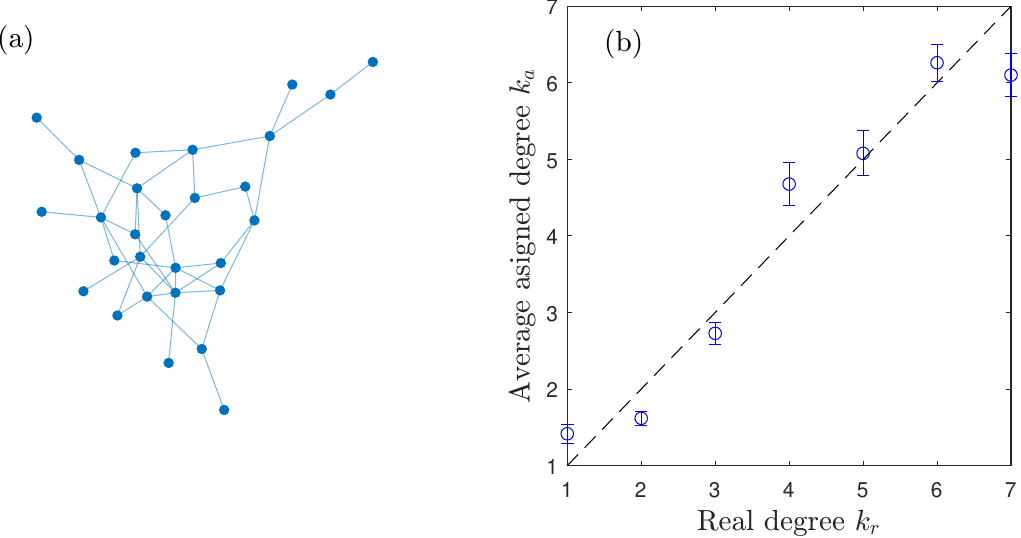}
    \caption{Inference of the nodes' degree of networks of electronic circuits based on the $k$-class network permutation  entropy $\langle {\cal H}_{\rm T} \rangle_k$ of the timeseries reported in Ref.~\cite{Vera2020}. (a) Structure connectivity of the electronic circuit network used as a ground truth. (b) Average assigned degree $k_a$ versus the real degree $k_r$ obtained when using a single network as a training reference.}
    \label{fig5_exp}
\end{figure}

Therefore, these experimental datasets provide the ideal testbed for our inference method and to predict the circuits connectivity by means of the network permutation entropy of each timeseries' circuit. In order to do so, we choose a weak coupling condition (level 9 over 100) and, for only one of the network configurations [the one that is used as a ground truth reference, plotted in Fig. \ref{fig5_exp} (a)], we calculate the average $k$-class network permutation entropy $\langle {\cal H}_{\rm T} \rangle_k$.

The output of this calibration procedure is a function that maps the node degree classes of the network used as a ground truth and the corresponding assigned network permutation entropies. One possible way is to produce a piecewise function $k_a({\cal H}_{\rm T})$ such that the sequence of intervals are defined by interpolating the entropies  measured in the experiment used as calibration for the degrees  $k$ and $k+1$, that is, $T_H(k)=[\langle {\cal H}_{\rm T} \rangle_{k+1} - \langle {\cal H}_{\rm T} \rangle_{k}]/2$ for $k=1,\dots,k_{\rm max}-1$. Now, for each node $i$ in any network different from the one used as a reference, we blindly assign a degree $k_a$ as a function of their dynamics using the following map:
\begin{equation}
  k_a^i= \left\{
    \begin{array}{ll}
			1 & \mbox{if \;} {\cal H}_{\rm T}^i<T_H(1)\\
                k & \mbox{if \; }  T_H(k)<{\cal H}_{\rm T}^i<T_H(k+1) \mbox{; \;\;  } k=2,\dots,k_{\rm max}-1\\
     		k_{\rm max} & \mbox{if \;} {\cal H}_{\rm T}^i>T_H(k_{\rm max}-1)  
       \end{array} \right.
\end{equation}

Since the real degree $k_r^i$ of the node $i$ is available in the dataset, we can compare the predicted value $k_a^i$ with the real one. In Fig. \ref{fig5_exp}(b), we plot the assigned degree versus the real one averaged for all the nodes in the 19 networks. Notice that these networks are very small and relatively sparse, with a maximum degree  $k_{\rm max}=7$ and, therefore, the degree sequence spans  a much narrower interval than in the SF networks used in the simulations shown in Fig. \ref{fig4}. Remarkably, even in this degree-constrained scenario and despite the noise inherent to an experimental environment, we obtain that for the $91\%$ of the nodes $|k_r^i-k_a^i| \le 1$, constituting a very high confidence level.

\section{Conclusions} 

Ordinal measures provide a valuable collection of tools for analyzing correlated data series. However, the use of these methods to understand information interchange in coupled networks and the interaction between dynamics and structure during the synchronization process remains relatively unexplored. In this study, using networks of coupled R\"ossler systems in chaotic regime, we assessed the performance of the standard ordinal permutation entropy ${\cal H}_0$ compared to the network permutation entropy ${\cal H}_T$, which captures information about transitions between ordinal patterns, and applied the proposed methodology to infer the connectivity of experimental datasets of networks of nonlinear circuits.

Whereas there exist other measures, such as statistical permutation complexity \cite{Tlaie2019} and ordinal structurality \cite{Letellier2020}, which have demonstrated their usefulness as proxies for degree distributions, our findings highlight the ordinal transition entropy as a more effective method for distinguishing topological roles, and producing more satisfactory outcomes, particularly for lower embedding dimensions.

Many methods focused on the structure-function relationship are primarily intended to infer the detailed connectivity network, down to the level of the individual links, from time series. However, in many cases, knowledge of centrality roles alone is sufficient for designing successful interventions in the dynamics. Therefore, we anticipate that our results, which do not rely on pairwise correlations between timeseries, will be of particular interest in the context of functional networks and other scenarios in which the underlying structural information is inaccessible.

\acknowledgments This research was supported by the Spanish  Ministerio de Ciencia e Innovación under Project PID2020-113737GB-I00.

\bibliography{references}
\end{document}